\documentclass{article}
\usepackage{amsfonts}
\usepackage{amsmath,amsthm}
\usepackage{graphicx,epstopdf,epsfig,multirow,epic,bm}
\usepackage{color}
\usepackage{multicol}
\usepackage{CJK}

\normalsize \rm
\begin{document}

\begin{center}
{\Large  \textbf { An ensemble of random graphs with identical degree distribution }}\\[12pt]
{\large Fei Ma$^{a,}$\footnote{~The author's E-mail: mafei123987@163.com. },\quad Xiaomin Wang$^{a,}$\footnote{~The author's E-mail: wmxwm0616@163.com.} \quad  and  \quad  Ping Wang$^{b,c,d,}$\footnote{~The corresponding author's E-mail: pwang@pku.edu.cn.} }\\[6pt]
{\footnotesize $^{a}$ School of Electronics Engineering and Computer Science, Peking University, Beijing 100871, China\\
$^{b}$ School of Software and Microelectronics, Peking University, Beijing  102600, China\\
$^{c}$ National Engineering Research Center for Software Engineering, Peking University, Beijing, China\\
$^{d}$ Key Laboratory of High Confidence Software Technologies (PKU), Ministry of Education, Beijing, China}\\[12pt]
\end{center}

\begin{quote}
\textbf{Abstract:} Degree distribution, or equivalently called degree sequence, has been commonly used to be one of most significant measures for studying a large number of complex networks with which some well-known results have been obtained. By contrast, in this paper, we report a fact that two arbitrarily chosen networks with identical degree distribution can have completely different other topological structure, such as diameter, spanning trees number, pearson correlation coefficient, and so forth. Besides that, for a given degree distribution (as power-law distribution with exponent $\gamma=3$ discussed here), it is reasonable to ask how many network models with such a constraint we can have. To this end, we generate an ensemble of this kind of random graphs with $P(k)\sim k^{-\gamma}$ ($\gamma=3$), denoted as graph space $\mathcal{N}(p,q,t)$ where probability parameters $p$ and $q$ hold on $p+q=1$, and indirectly show the cardinality of $\mathcal{N}(p,q,t)$ seems to be large enough in the thermodynamics limit, i.e., $N\rightarrow\infty$, by varying values of $p$ and $q$. From the theoretical point of view, given an ultrasmall constant $p_{c}$, perhaps only graph model $N(1,0,t)$ is small-world and other are not in terms of diameter. And then, we study spanning trees number on two deterministic graph models and obtain both upper bound and lower bound for other members. Meanwhile, for arbitrary $p(\neq1)$, we prove that graph model $N(p,q,t)$ does go through two phase transitions over time, i.e., starting by non-assortative pattern and then suddenly going into disassortative region, and gradually converging to initial place (non-assortative point). Among of them, one ¡°null¡± graph model is built.


\end{quote}

\vskip 1cm

\section{INTRODUCTION}

The last few decades have seen an increasing interest of studying complex networks in a variety of disciplines, including statistic physics, applied mathematics, theoretical computer science, biology as well chemistry, and so forth. In general, a network is a collection of discrete terms, named vertices (or nodes), connected by edges (or lines). The widely studied network models consist of the World Wide Web (WWW), the Internet, sexual contract network, scientist cooperation networks, friendship networks, protein interaction networks,
metabolic networks \cite{Watts-1998}-\cite{W-L-2018}. Different from classic random graph models with poisson distribution provided by Erd\"{o}s and R\`{e}nyi \cite{P-E-1959}, in some context, these real-life networks mentioned above appear to follow the highly skewed degree distribution, commonly known as power-law distribution in form. This type of distribution suggests that there are a small number of vertices possessing a large fraction of connections in a network, and vice versa.

Since then, degree distribution (degree sequence) can be regarded as a simple yet useful measure to answer whether a given network model is power-law or not. On the other hand, for a given degree distribution (degree sequence), how to rebuild a least one corresponding network model is of important interest and has been paid more attention. More generally speaking, it is not easy to produce a better model meeting a given degree sequence. Hence, some helpful techniques have been developed. Among of which, the popularly utilized method is generating function, $G(x)=\sum P_{i}x^{i}$, based on its own advantages, for instance (1) average degree $\langle k \rangle =G'(1)=\sum kP_{k}$, (2) probability  $P(k)=\frac{1}{k!}\frac{d^{k}G(x)}{dx^{k}}|_{x=1}$, and (3) higher moment $\langle k^{n} \rangle=\left[\left(x\frac{d}{dx}\right)^{n}G(x)\right]|_{x=1}$, and so on \cite{M. E. J. Newman-2005}. It is hesitated for ones to make only use of degree distribution to describe complicated networks. Put it another way, it is likely for some network models with identical degree distribution to behave different topological structure among them. Here, to further concrete this assertion, we build a graph space $\mathcal{N}(p,q,t)$ where probability parameters $p$ and $q$ hold on $p+q=1$ in which each member obeys a unique degree distribution $P(k)\sim k^{-\gamma}$ with $\gamma=3$. As we will show shortly, there in practice are some pronounced differences among all the members in graph space $\mathcal{N}(p,q,t)$. For example, it is convenient for one to achieve a transformation of a small-world member into a larger one by just tuning probability parameter $p$. For each time step $t$, graph model $N(1,0,t)$ always has pearson correlation coefficient $r=0$ in comparison with other different members of graph space $\mathcal{N}(p,q,t)$ in the thermodynamic limit.
Different topological structures might lead to some non-similar structure features. Among of various of features, we here just focus on spanning trees number and obtain an  inequality relevant to spanning trees numbers of all the members of graph space $\mathcal{N}(p,q,t)$.

As said in Ref \cite{M. E. J. Newman-2003}, consider a network, its pearson correlation coefficient $r$ must belong to a range between $-1$ and $1$ and positive value $r$ represents this network has assortative mixing, negative value $r$ for disassortatively connected network, and value $0$ for intermediary. However, during a finite number of time steps in question, the ``null" graph model $N(1,0,t)$ seem to disprove the above statement because although so clear to see vertices with great degree connected by a number of small degree vertices, its pearson correlation coefficient are a unchanged constant, namely $r=0$. Maybe our finding can help one to understand the relation between pearson correlation coefficient $r$ and network structure properties deeply.

This paper can be organized by the following several Sections. In Section 2, in order to successfully build our graph space $\mathcal{N}(p,q,t)$ whose each member behaves a designed degree distribution $P(k)\sim k^{-\gamma}$ with $\gamma=3$, we introduce two types of operations, type-A operation and type-B operation. After that, Section 3 is made up of three subsections in which we mainly discuss diameter, spanning trees number and pearson correlation coefficient, respectively. For the concrete outline of this paper, we have to close this paper by making an elaborated conclusion and bringing some discussions associated with our graph space $\mathcal{N}(p,q,t)$ for future work in the last section.

\section{OPERATIONS AND CONSTRUCTION}

In this section, we will generate a class of random graphs with identical degree distribution which can be used to span a graph space, denoted by $\mathcal{N}(p,q,t)$. Here, both probability parameters $p$ and $q$ hold $p+q=1$ with $0\leq p,q\leq1$, and $t$ represents time step. Put it another way, for two arbitrary members $N(p_{i},q_{i},t)$ and $N(p_{j},q_{j},t)$ at random chosen from graph space $\mathcal{N}(p,q,t)$, they must have same ¡°degree sequence¡± as each other. To do this, we need to introduce two well-studied operations, called in this paper type-A operation and type-B operation respectively, which are described in more detail, as follows

\textbf{Type-A operation} For a given edge $uv$ with two vertices $u$ and $v$, bringing an edge $xy$ on vertices $x$ and $y$ and then connecting vertex $u$ with $x$ and $v$ with $y$ using two new edges, respectively, produces a cycle $C_{4}$. Such a process is Type-A operation, also defined as rectangle method in our previous works \cite{M-f-2019}, shown in Fig.1(a).

\textbf{Type-B operation} For a given active edge $uv$ with two vertices $u$ and $v$, bringing two vertices $x$ and $y$, connecting vertex $x$ with two endpoints of edge $uv$ by two new edges and similar connections for vertex $y$ and vertex pair $u$ and $v$, as well deleting active edge $uv$, together produces a cycle $C_{4}$. Such a process is Type-B operation, also commonly defined as fractal method or as diamond method, seeing Fig.1(b).

\begin{figure}
\centering
  \includegraphics[height=7cm]{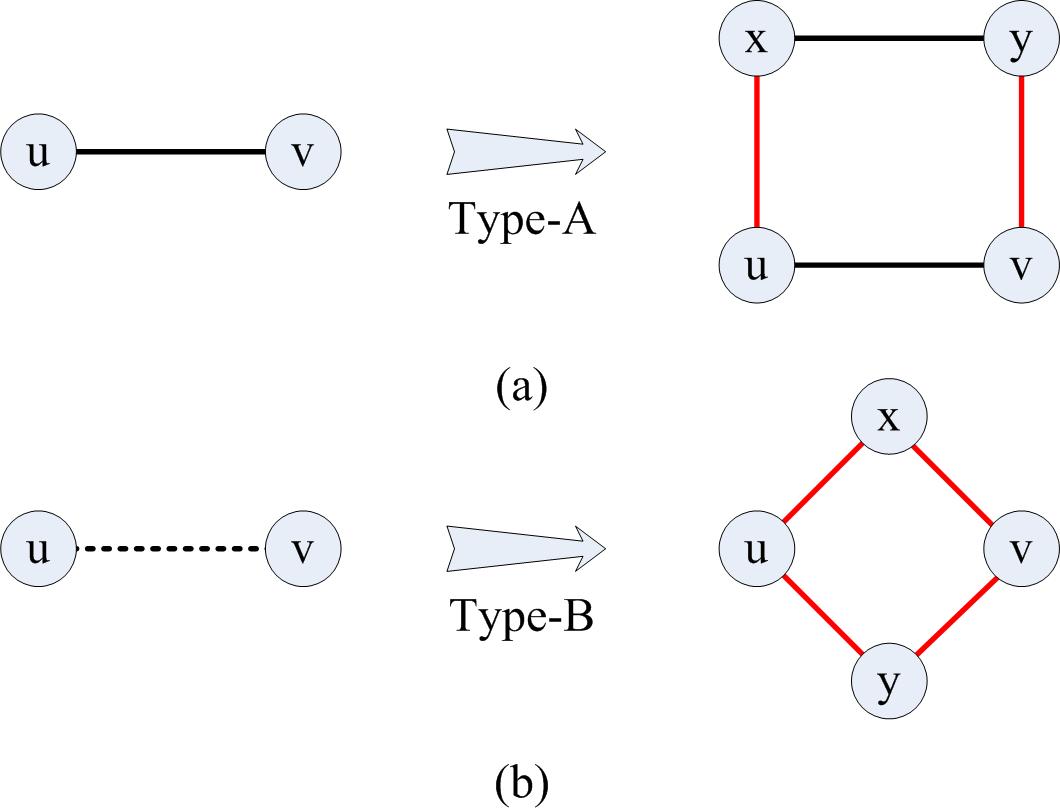}\\
{\small Fig.1. The diagram of both type-A and type-B operations.  }
\end{figure}

It in reality is not necessary to distinguish the two cycles $C_{4}$ obtained from both operations introduced above from the theoretical point of view. Nonetheless, we in this paper prefer to refer to the former cycle $C_{4}$ as a rectangle and the later as a diamond. Only reason is able to organize the outline of our work conveniently. Isomorphism between them in structure can not guarantee identical function taking place on them as we will depict shortly. Put this in mind and keep on following.

There are two popular manners in which most of pre-existing graph models can be built up. One is to first construct graphs using some rules except for degree sequence (degree distribution) and next to study their own properties including degree sequence. Such well-defined examples have scale-free BA-model due to \cite{Albert-1999-1}, small-word WS-model by \cite{Watts-1998},
Model of Apollonian Networks \cite{N-Z-20O8}, sierpinski networks \cite{N-I-2017}. The other is, for a given degree sequence, to establish an available graph consistency with that designed degree sequence. From their appearance, the two process above may be thought of being inverse. The later in theory is much difficult than the former. So far, some useful algorithms and methods have been provided to do so. Included studies have generating function \cite{M. E. J. Newman-2003} and algorithms \cite{m-m-1998}. We attempt to construct our graph space $\mathcal{N}(p,q,t)$ with identical degree distribution in the later manner based on a given degree sequence. Degree of a vertex is the number of vertices which it is connected, denoted by $k$.

As shown in most natural and man-made complex networks, small-world property and scale-free feature are highly prevalent. In order to protray such two characters and to study function occurring over networks of this type, a large number of models have be generated. For our purpose, this paper focuses mainly on a graph space $\mathcal{N}(p,q,t)$ whose each member follows a given power-law degree distribution in form

$$P(k)\sim k^{-\gamma}$$
where symbol $P(k)$ represents a probability for  choosing at random a vertex with degree equal to $k$ from the entire network. Taking into account the background work done by Barabasi and Albert making using of the mean-field theory, we make an equivalent assumption that
let power-law parameter $\gamma$ be equal to $3$ not to anything else. In general, it is not easy to reconstruct a graph obeying power-law degree distribution with a given value $\gamma$. A widely used technique is generating function

$$G(x)=\sum P_{i}x^{i}.$$
There are an increasing deal of researches relevant to constructing graphs utilizing generating function, refer to \cite{M. E. J. Newman-2003}.

Taking useful advantage of two types of operations and generating function, let us now turn our sight into building up graph space $\mathcal{N}(p,q,t)$.

\begin{figure}
\centering
  \includegraphics[height=7cm]{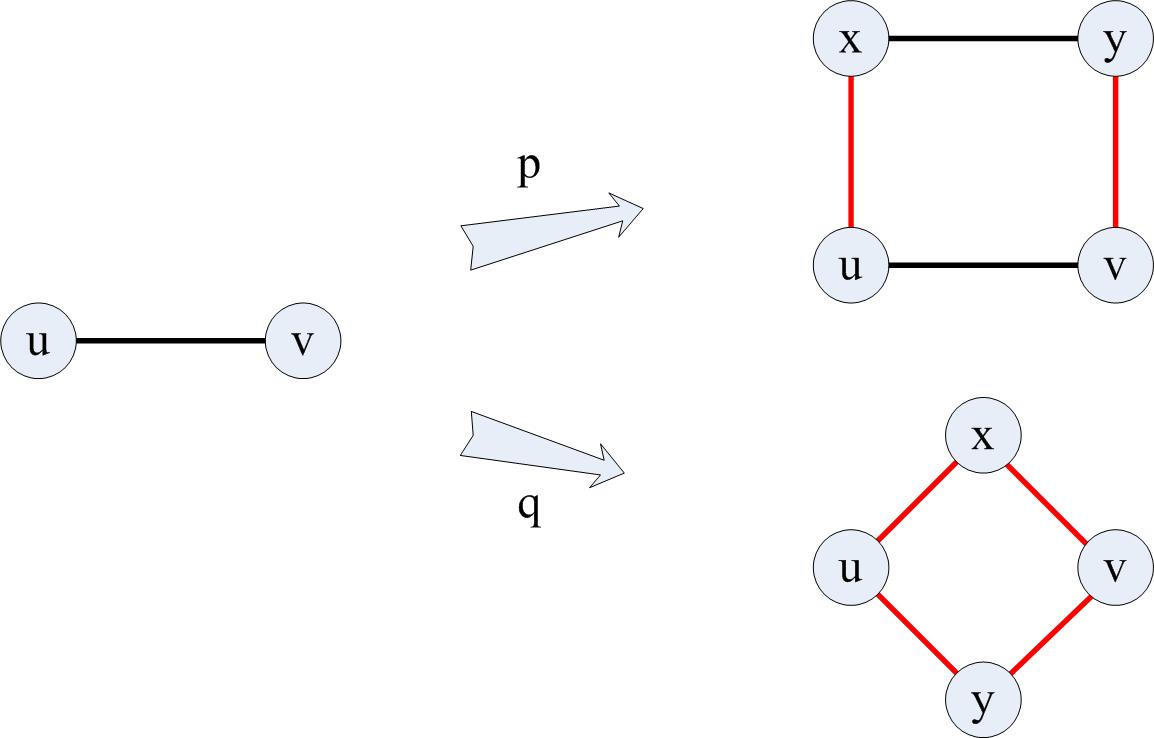}\\
{\small Fig.2. The diagram of operation implemented on an edge. In the process of constructing $N(p,q,t)$ from $N(p,q,t-1)$ ($t\geq1$), a probability $p$ for each edge of $N(p,q,t-1)$ to apply type-A operation and complementary probability $q$ for applying type-B operation.      }
\end{figure}

\textbf{CONSTRUCTION}

First of all, the seminal graph $N(0)$ is a cycle $C_{4}$. To introduce various graphs into graph space $\mathcal{N}(p,q,t)$, we employ randomization method, namely using a pair of probability parameters $p$ and $q$ satisfying $p+q=1, 0\leq p, q\leq1$. Thus, in order to obtain the next graph $N(p,q,t)$ from $N(p,q,t-1)$ ($t\geq1$), one just need to apply type-A operation to each edge of graph $N(p,q,t-1)$ with probability $p$ or apply type-B operation to each edge of graph $N(p,q,t-1)$ with complementary probability $q$, shown in Fig.2. As studied in previous literatures \cite{Z-Z-Z-2009}, our graph space $\mathcal{N}(p,q,t)$, after $t$ time steps,  will consist of a single element at $p=0$ or $p=1$ and becomes a deterministic graph whose some topological properties of interest have be discussed in detail. To make this paper much self-contained, we simply list a few common properties shared by both deterministic graphs $N(1,0,t)$ and $N(0,1,t)$ as follows. Graphs $N(1,0,t)$ and $N(0,1,t)$ both have same vertex number (order),  $|V(1,0,t)|=|V(0,1,t)|=\frac{8\cdot4^{t}+4}{3}$, and identical edge number (size),  $|E(1,0,t)|=|E(0,1,t)|=4^{t+1}$. And then, they are sparse in terms of a small average degree $\langle k\rangle$ close to $3$. With the help of evolution process of two graphs described above, it can be easily seen that they also follow a unique degree distribution $P(k)=k^{-\gamma}$ with a constant exponent $\gamma=3$ as assumed by us previously. It is worth noting that both sparsity and power-law degree distribution can in general be found in variety of complex networks around us. If one only concerns on the two features, then graphs $N(1,0,t)$ and $N(0,1,t)$ appear to be reasonable candidates. On the other hand, as shown in recent researches, most networks have a great degree of transitivity or clustering, i.e., there is a high probability that
¡°two of you friends are also friends of one another¡±. Unfortunately, graph models $N(1,0,t)$ and $N(0,1,t)$ have no clustering by virtue of nonexistence of triangle. In order to better mimic real-life complex networks, graph families with tunable clustering are being discussed in our another paper \cite{W-X-M-2019}. Here we aim at studying graphs with identical degree distribution and some of their applications in topological terms.

More generally, each member of our graph space $\mathcal{N}(p,q,t)$ obeys similar power-law degree distribution to the proceeding two deterministic models $N(1,0,t)$ and $N(0,1,t)$. One of most important reasons for this is that the generation model using type-A operation has a topological structure with that of type-B operation in common, i.e., resulting model is a cycle $C_{4}$. Clearly the two swap procedures preserve the degree sequence. Hence this probability parameter $p$ has no influence on degree distribution of each member of graph space $\mathcal{N}(p,q,t)$ but on other topological structure parameters over each member as we will show shortly. This is why we want to build graph space $\mathcal{N}(p,q,t)$ in which although having same degree distribution, all members are different from each other under other topological terms. Except for that, what's more, there is very likely for a given probability parameter $p$ and a provided time step $t$ to construct various $N(p,q,t)$ in structure. What we present here will be   significant constituents in the following sections.

\section{PROPERTIES AND DISCUSSIONS }

In the last few decades, there are an increasing number of literatures published to unveil intrinsic characters behind complex networks and to understand many functions of interest taking place over this types of networks. Among of them, the scale-free feature and small-world property are two prominent findings. As mentioned in Section 2, each member of  graph space $\mathcal{N}(p,q,t)$ follows a unique power-law degree distribution and displays scale-free feature with respect to two vital mechanisms introduced by Barabasi and Albert. However small-word property describes another apparent phenomena that in most case, there seems to be many connections of short length linking two arbitrary persons chosen randomly from our living world. Nowadays, this interesting fact has be called the six-degree separation theory and accepted by people. Mathematically, one might find out a few short length connections of a source vertex to an arbitrary vertex in a connected complex network modeling real-life relationships among people in society. The length of such a connection between a pair of vertices is usually defined as distance. From the graph theory point of view, the diameter $D$ is regarded as the maximum of all distances between any two vertices in a connected and undirected graph. Diameter is itself a feature of a graph topology and can be simply used to measure information delay over a network. Particularly,  in information science, diameter $D$ suggests a transmission efficiency of information over the whole network. More usually, the greater diameter $D$ is, the poorer transmission efficiency is. Therefore, let us first study diameter both analytically and experimentally.

\textbf{DIAMETER}

Before proceeding, let us focus on two simplest cases, namely deriving exact expressions to diameters of both deterministic graph models $N(1,0,t)$ and $N(0,1,t)$, respectively.

Considering, in the process of generating $N(1,0,t)$ by $N(1,0,t-1)$, only applying type-A operation to each edge of $N(1,0,t-1)$, the diameter $D(1,0,t)$ of graph model $N(1,0,t)$ must be spanned by virtue of graph model $N(1,0,t-1)$ own diameter $D(1,0,t-1)$. By definition of diameter, take a path with length precisely equal to  $D(1,0,t-1)$ from graph model $N(1,0,t-1)$ and denote it by $u_{1}u_{2},...,u_{D(1,0,t-1)+1}$. After that, based on producing $N(1,0,t)$, we make using of type-A operation on each edge $u_{i}u_{i+1}$ ($1 \leq i\leq D(1,0,t-1)$) and obtain a length $D(1,0,t-1)+2$ path, similarly for arbitrary path of various length in graph model $N(1,0,t-1)$. So we may write a relationship between $D(1,0,t)$ and $D(1,0,t-1)$, i.e., $D(1,0,t)=D(1,0,t-1)+2$. With the condition $D(1,0,0)=2$, one immediately captures a closed-form solution to diameter $D(1,0,t)$, that is, $D(1,0,t)=2(t+1)$. Armed with a fact $\ln|V(1,0,t)|\sim\ln4^{t+1}=(t+1)\ln4$, we find out a connection between diameter $D(1,0,t)$ and vertex number $|V(1,0,t)|$ according to $D(1,0,t)\sim\ln|V(1,0,t)|$. In other words, in the limit of large graph size, the diameter of $N(1,0,t)$ is much smaller than its own order and only scales logarithmically. Such a phenomena can be easily found in a great number of real-life networks, indicating that graph model $N(1,0,t)$ is like such real-life networks and has small-world feature.

By analogy with the development of $D(0,1,t)$, one can write a recursive connection of diameter $D(0,1,t)$ in graph model $N(0,1,t)$ to $D(0,1,t-1)$ on $N(0,1,t)$ as
\begin{equation}\label{Diameter-1}
D(0,1,t)=2D(0,1,t-1).
\end{equation}
Compared to type-A operation, the function of type-B operation is to replace any edge $uv$ by a length $2$ path which does account for the prefactor  $2$ in the right-hand side of Eq.(\ref{Diameter-1}). Plugging this initial value $D(0,1,0)=2$ into Eq.(\ref{Diameter-1}) yeilds a solution to diameter $D(0,1,t)$, $D(0,1,t)=2^{t+1}$. Different from that case of graph model $N(1,0,t)$, diameter $D(0,1,t)$ is not approximately equal to $\ln|V(0,1,t)|$ but to a square root value of the vertex number of graph model $N(0,1,t)$, directly indicating $N(0,1,t)$ is large-scale.

So far, we in practice study two specified cases and our discussions about diameters of other members of graph space $\mathcal{N}(p,q,t)$ still remain unanswered thoroughly. One of most important reasons for this is to introduce probability parameter $p$ into the generation process of graph. Even though for given parameters $p$ and $t$, there may be a lot of generation graphs $N(p,q,t)$ with various topological structure parameters including diameter except for degree distribution. So we on computer manipulate simulation for the dynamical tendency of diameter over time. Due to space and time memory, we make a reasonable constraint, setting $0\leq t\leq 10$ and $0\leq p \leq 1$. The resulting diagram is illustrated in Fig.3.

\begin{figure}
\centering
  \includegraphics[height=7cm]{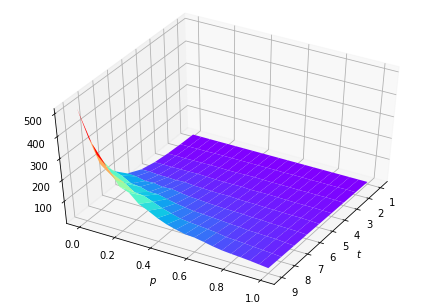}\\
{\small Fig.3. The diagram of diameter with constraints,  $0\leq t\leq 10$ and $0\leq p \leq 1$, of graph space $\mathcal{N}(p,q,t)$.   }
\end{figure}

As is clear from Fig.3, the diameter averaged over graph space $\mathcal{N}(p,q,t)$ as a whole sharply diverges with decreasing value $p$. In order to see clearly the influence from probability parameter $p$ on diameter, we give three specified values with $p$, $p=0.1, 0.5, 0.9$, and then run our algorithm for each value $p$ exactly $10$ times, see Fig.4. With the help of the three diagrams, we attempt to capture an appropriately analytical solution to diameter. Different from the previous calculations for both deterministic graph models, we need to consider three contributions into the change of diameter in the process of obtaining graph model $N(p,q,t)$ from $N(p,q,t-1)$ by implementing both type-A and type-B operations in question. We first choose at random an arbitrary path of length equivalent to $D(p,q,t-1)$, denoted by $u_{1}u_{2},...,u_{D(1,0,t-1)+1}$. Such a choice is reasonable due to our assumption that each edge is taken use of type-A or type-B operations independently. $\emph{Case 1.}$ If one applies type-A operation to the two edges $u_{1}u_{2}$ and $u_{D(1,0,t-1)}u_{D(1,0,t-1)+1}$ simultaneously, then $D(p,q,t)=pD(p,q,t-1)+2+2(1-p)D(p,q,t-1)$. $\emph{Case 2.}$ Only one of the two edges $u_{1}u_{2}$ and $u_{D(1,0,t-1)}u_{D(1,0,t-1)+1}$ is manipulated by type-A operation, then $D(p,q,t)=pD(p,q,t-1)+1+2(1-p)D(p,q,t-1)$. $\emph{Case 3.}$ Neither edge $u_{1}u_{2}$ nor edge $u_{D(1,0,t-1)}u_{D(1,0,t-1)+1}$ are not changed using type-A operation and hence one can write $D(p,q,t)=pD(p,q,t-1)+2(1-p)D(p,q,t-1)$. On the one hand, fortunately, we always encounter case 3 at each time step and can obtain a recursive equation to diameter, $D(p,q,t)=(2-p)^{t}D(p,q,0)$. On the other hand, we  may derive another expression for  diameter, namely $D(p,q,t)=(2-p)^{t}(D(p,q,0)+\lambda)-\lambda$, under the situation of case 1. Here $D(p,q,0)=2$ and $\lambda=2/(1-p)$. Due to $p$ lying in range $0<p<1$, either of the two approximate solutions above indicates that diameter grows exponentially with increasing time $t$. To show this, we in appendix A provide two diagrams.

\begin{figure}
\centering
  \includegraphics[height=3.8cm]{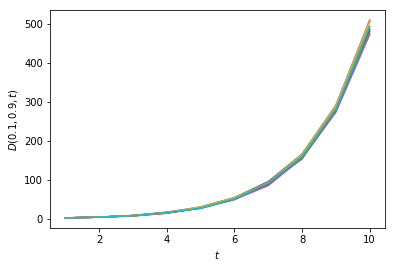}\quad
  \includegraphics[height=3.8cm]{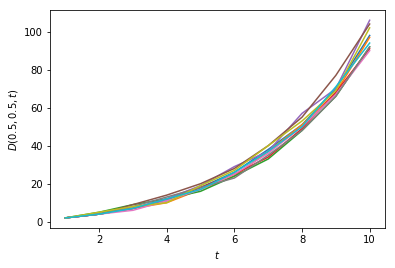}\quad
  \includegraphics[height=3.8cm]{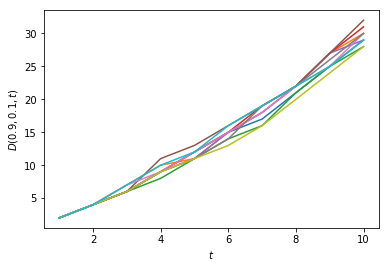}\\
{\small Fig.4. The diagrams of diameters averaged over graph space $\mathcal{N}(p,q,t)$ given $p=0.1$, $p=0.5$ as well $p=0.9$, respectively. It is clear for the eye that curvature of the curve represents the degree of influence from probability parameter $p$. Each colored line hints a result by running our algorithm exactly once.   }
\end{figure}

Compared to two extreme cases, $p=1$ and $p=0$, for arbitrary probability parameter $p$ meeting $0<p <1$, there appears to be $D(1,0,t)\leq D(p,q,t)\leq D(0,1,t)$. It's too early to make such a conclusion. It is not hard to find out a counterpart, refer to Appendix A for detail. However, with some additional conditions, such as the limit of large graph size, we indeed arrive at the  following proposition about both upper bound and lower bound of the diameter of each member belonging to graph space $\mathcal{N}(p,q,t)$.

\textbf{Proposition 1}
For any member $N(p,q,t)$ of network space $\mathcal{N}(p,q,t)$ with $p_{c}<p \leq1$ where threshold value $p_{c}$ is larger enough than $1/4^{t}$, its diameter $D(p,q,t)$ must follow the following inequality in the limit of large graph size

\begin{equation}\label{Diameter-2}
D(1,0,t)\leq D(p,q,t)\leq D(0,1,t).
\end{equation}

From Eq.(\ref{Diameter-2}), it is worth noting that for a given parameter $0<\varepsilon(t)\ll1$ with the condition $\varepsilon(t)t\rightarrow\infty$, only if two distinct probability parameters $p_{i}$ and $p_{j}$ meet a inequality $\varepsilon(t)\leq|p_{i}-p_{j}|$, then in the thermodynamic limit the ratio $D(p_{i},q_{i},t)$ and $D(p_{j},q_{j},t)$ will be either infinite or zero according to importance limit $\left(1+
\frac{1}{t}\right)^{t}$, indicating a smaller change on parameter $p$ will make two graph models considerably different, at least in this case of diameter.

In a words, for an arbitrary member of network space $\mathcal{N}(p,q,t)$, Type-B operation can make diameter  change more severely than Type-A operator by both analytically and experimentally. By varying value $p$ from $1$ to $0$, the small-world graph model $N(p,q,t)$ is  abruptly transformed into another type of graphs whose scale is larger. As said in section 2, degree distribution can not play a role to better distinguish each member of graph space $\mathcal{N}(p,q,t)$. By our discussion  here, diameter seems to be an available measure used to do so. In addition, there are some helpful indices for understanding the difference among graphs with identical degree distribution. By contrast, the choice of other types of measures is a matter of convenience. The remainder of this section is to introduce the other two topological structure parameters, spanning trees number and pearson correlation coefficient.

\textbf{SPANNING TREES}

Spanning trees number is always considered as an important structure invariant relevant to several kinds of dynamical functions on networks, for instance reliability \cite{G-J-S-2003}, synchronization capability \cite{N-T-A-E-2006}, random walks \cite{P-M-2000}-\cite{Z-Z-2013}, to name just a few. Hence, in the past most attentions have been focused on enumerating the number of spanning trees of special network models. As before, let us put our sight into two particular cases, $N(1,0,t)$ and $N(0,1,t)$.

According to our previous results in Ref \cite{F-M-20181} , for deterministic graph model $N(1,0,t)$, we can write

\begin{equation}\label{Spanning tree-1}
\left\{\begin{aligned}
    &S(1,0,t)=4S^{3}(1,0,t-1)F(1,0,t-1)\\
    &F(1,0,t)=3S^{2}(1,0,t-1)F^{2}(1,0,t-1)
\end{aligned}
\right.
\end{equation}
where $S(1,0,t)$ is spanning trees number and $F(1,0,t)$ the number of \emph{2-(u; v)-forest}, refer to Ref \cite{F-M-20181}.

With the help of some simpler arithmetic, one can obtain an closed-form solution to spanning trees number as follows

\begin{equation}\label{Spanning tree-2}
S(1,0,t)=3^{\frac{4^{t+1}-1}{3}}\times\left(\frac{4}{3}\right)^{\frac{4^{t+1}+6t+5}{9}}=3^{\psi_{1}(1,0,t)}
\times4^{\psi_{2}(1,0,t)}.
\end{equation}

Analogously, a group of equations between $S(0,1,t)$ and $F(0,1,t)$ can be read
\begin{equation}\label{Spanning tree-3}
\left\{\begin{aligned}
    &S(0,1,t)=4S^{3}(0,1,t-1)F(0,1,t-1)\\
    &F(0,1,t)=4S^{2}(0,1,t-1)F^{2}(0,1,t-1)
\end{aligned}
\right.
\end{equation}

Under the initial condition $S(0,1,0)/F(0,1,0)=4/4=1$, an exact expression for $S(0,1,t)$ is

\begin{equation}\label{Spanning tree-4}
S(0,1,t)=4^{\frac{4^{t+1}-1}{3}}=4^{\psi(0,1,t)}.
\end{equation}

Apparently, graph models $N(1,0,t)$ and $N(0,1,t)$ both have a unique degree distribution but possess distinct spanning trees number. In such a situation, degree sequence can not be used as a reliable index to distinguish them but spanning trees number seems to be a better replacement.

There appears to exist a connection of $S(1,0,t)$ to $S(0,1,t)$ because of their underlying graphs both having identical degree distribution. Indeed, based on Eqs.(\ref{Spanning tree-2}) and (\ref{Spanning tree-4}), one can easily see the following equation

\begin{equation}\label{Spanning tree-5}
S(0,1,t)/S(1,0,t)=(4/3)^{2\frac{4^{t+1}-3t-4}{9}}, \quad and \quad \psi(0,1,t)=\psi_{1}(1,0,t)+\psi_{2}(1,0,t) .
\end{equation}
Eq.(\ref{Spanning tree-5}) provides a strong proof to a fact that for an arbitrary member of network space $\mathcal{N}(p,q,t)$, Type-B operation can produce more spanning trees than Type-A operation and hence plays a most significant role in the process of constructing network models. Intuitively, consider both $D(0,1,t)/D(1,0,t)$ and $S(0,1,t)/S(1,0,t)$, we might state that spanning trees  number is a better measure distinguishing each member of graph space $\mathcal{N}(p,q,t)$ than diameter.

Similarily, we immediately arrive at the second proposition about both upper bound and lower bound of the total numbers of spanning trees of each member belonging to network space $\mathcal{N}(p,q,t)$.

\textbf{Proposition 2} For any member $N(p,q,t)$ of  network space $\mathcal{N}(p,q,t)$, its spanning trees number must satisfy the following inequality

\begin{equation}\label{Spanning tree-6}
S(1,0,t)\leq S(p,q,t)\leq S(0,1,t).
\end{equation}

As known, given a pair of parameters $p$ and $q$, there will be a great number of candidates being in network space $\mathcal{N}(p,q,t)$. Hence we are able to assert that the coordinate $|\mathcal{N}(p,q,t)|$ of network space $\mathcal{N}(p,q,t)$
should be too large in the limit of large generation $t$ to enumerate them even utilizing current computers. The ratio described in Eq.(\ref{Spanning tree-5}) however is smaller than the coordinate $|N(p,q,t)|$, we are convinced that there must exist two distinct members $N(p_{1},q_{1},t)$ and $N(p_{2},q_{2},t)$ with identical spanning trees number. If so, spanning trees number will not be adequate to differentiate some members of network space $\mathcal{N}(p,q,t)$. Other useful indices should be adopted to do this. Therefore looking for other new measures will an interesting research topic in the neat future and is also one of our present focuses.

\textbf{PEARSON CORRELATION COEFFICIENT}

To further put forward our task to better distinguish difference among members of network space $\mathcal{N}(p,q,t)$, this subsection mainly focuses on another topological structure parameter, the correlations between properties of adjacent vertices,  known as pearson correlation coefficient $r$ \cite{M-E-J-Newman2003}.

Recent studies have proven that for many real-world networks, the degrees of vertices at either endpoint of an edge chosen randomly are not independent, but are
correlated with one another. With the help of pearson correlation coefficient $r$, all most of social networks are turn out to have positive value $r$ and hence are assortatively constructed. On the other hand,  non-social networks, such as technological and Biological networks, have disassortative mixing pattern. For convenience and our purpose, below is a brief introduction to knowledge relevant to pearson correlation coefficient $r$ \cite{M-E-J-Newman2003}.

More commonly, a normalized assortativity coefficient can be obtained, as follows:
\begin{equation}\label{Pearson-1}
r=\frac{1}{\delta^{2}_{q}}\sum\limits_{j,k}jk(e_{jk}-q_{j}q_{k})
\end{equation}
here $e_{jk}$ is the fraction of edges running vertices of degree $j$ and $k$ in a network,  $q_{k}$ is equal to $(k+1)P(k+1)/\sum_{k} kP(k)$ and $\delta_{q}$ is the standard deviation of the distribution $q_{k}$. For brevity, Eq.(\ref{Pearson-1}) is usually translated into the next form

\begin{equation}\label{Pearson-2}
r=\frac{|E(p,q,t)|^{-1}\sum\limits_{e_{ij}\in E(p,q,t)} k_{i}k_{j}-\left[|E(p,q,t)|^{-1}\sum\limits_{e_{ij}\in E(p,q,t)} \frac{1}{2}(k_{i}+k_{j})\right]^{2}}{|E(p,q,t)|^{-1}\sum\limits_{e_{ij}\in E(p,q,t)} \frac{1}{2}(k^{2}_{i}+k^{2}_{j})-\left[|E(p,q,t)|^{-1}\sum\limits_{e_{ij}\in E(p,q,t)} \frac{1}{2}(k_{i}+k_{j})\right]^{2}}
\end{equation}
suggesting that $r$ seems to satisfy $-1\leq r\leq 1$.

Along the same research line as the two subsection above, we first study graph models $N(1,0,t)$ and $N(0,1,t)$ due to their own deterministic structure. For $N(1,0,t)$, we can group all edges into $(t+1)(t+2)/2$  classes in a straightforward manner. Similarly, the total edges of graph model $N(0,1,t)$ ($t>0$) can be classified into $t$ families. Taking such classifications into algorithm for computing $r$ yields an illustration plotted in Fig.6. For probability parameter $p$ being in this range between $0$ and $1$, we also run algorithm and then obtain the next three panels from three different viewpoints, see Fig.7. With the help of both Fig.6 and Fig.7, we may say that for the first several time steps, parameter $p$ can have  considerable influence on pearson correlation coefficient $r(p,q,t)$ and makes the growth tendency of  $r(p,q,t)$ non-monotonous but fluctuant. In the thermodynamic limit,  $r(p,q,t)$ will become more and more smooth and ultimately trends to critical value zero. (As a guide for the eye, three specified pearson correlation coefficients are plotted in Appendix B.)
By numerical simulations and theoretical analysis, proposition 3 is vaild.

\begin{figure}
\centering
  \includegraphics[height=3.8cm]{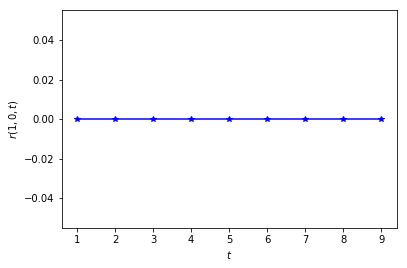}\quad
  \includegraphics[height=3.8cm]{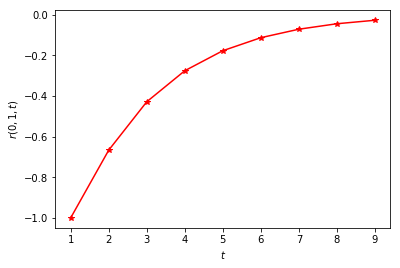}\\
{\small Fig.6. The diagrams of pearson correlation coefficients of graph models $N(1,0,t)$ and $N(0,1,t)$, $1\leq t\leq9$. Numerical simulations are in strong agreement with our analytical results.   }
\end{figure}

\textbf{Proposition 3}
For any member $N(p,q,t)$ of  network space $\mathcal{N}(p,q,t)$, in the limit of large graph size, its pearson correlation coefficient $r(p,q,t)$ must follow the following inequality

\begin{equation}\label{Pearson-5}
r(0,1,t)\leq r(p,q,t)\leq r(1,0,t).
\end{equation}

Similar discussions are suitable for pearson correlation coefficient $r(p,q,t)$. Although all the members of our network space $\mathcal{N}(p,q,t)$ have a unique power-law degree with exponent $\gamma=3$, they still display different behavior in this case of pearson correlation coefficient.

\begin{figure}
\centering
  \includegraphics[height=3cm]{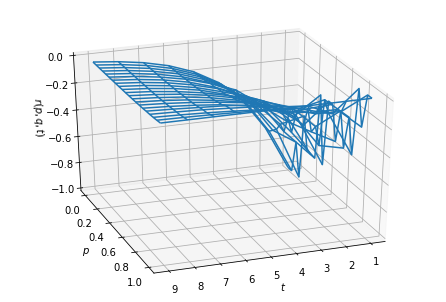}\quad
  \includegraphics[height=3cm]{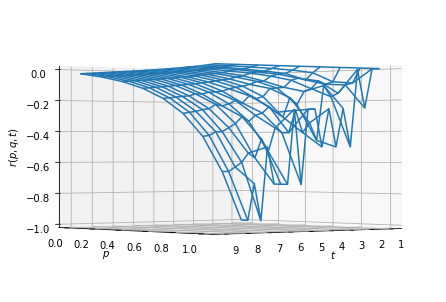}\quad
  \includegraphics[height=3cm]{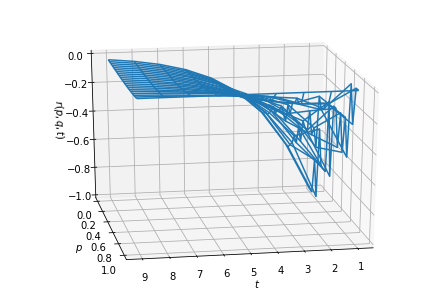}\\
{\small Fig.7. The diagrams of pearson correlation coefficient of each member of graph space $\emph{N}(p,q,t)$ with $1\leq t\leq9$ and $0\leq p\leq1$. }
\end{figure}

\section{CONCLUSION}

Recent studies of network structure have concentrated on a few of properties that appear to be popular in a great variety of complex networks. Among of them, the best studied are the scale-free feature and small-world property. Along such a research guide, to emphasis degree distribution in some cases can not perfectly distinguish many studied networks in question, we build a graph space $\mathcal{N}(p,q,t)$ using a given degree distribution whose each member follows power-law distribution with exponent $\gamma$ equivalent to a constant $3$. Under such a situation, there are three simple yet useful indices, namely, diameter, spanning trees number together with pearson correlation coefficient, adopted to better identify difference among each member of graph space $\mathcal{N}(p,q,t)$. By making a comparison in depth, we find out that this randomization parameter $p$ not only brings various types of graph models but also makes almost all graph models distinct with one another. By availably tuning value $p$, one can transform a small-world graph into an opposite one, further increase corresponding spanning trees number, simultaneously change assortative mixing over graph, i.e., from non-assortative mixing to disassortative pattern and ultimately return to assortative behavior.

Here, we want to state that network space $\mathcal{N}(p,q,t)$ studied is just an available example for showing, in some cases, degree distribution (degree sequence) is not well adequate to meet ones expectation. Hence, other helpful evaluating indicators are necessary and should be developed. Although network space $\mathcal{N}(p,q,t)$ has acted as a good role to imply some disadvantages of degree distribution, there exist a few flaws in its own structure. As known, cluster (or community) is prevailing in most of real-world networks, whereas no clustering phenomena is on our network space $\mathcal{N}(p,q,t)$ so that it is not suit to mimic real-world networks and to study dynamic function taking place on them. To accomplish a development of this model, we are doing so in our another manuscript which is out the scope of this paper. In addition,  of important interests will be discussed on network space $\mathcal{N}(p,q,t)$ in the next future, including random walks, synchronization, percolation.

\section*{ACKNOWLEDGEMENTS }
The research was supported by the National Key Research and Development Plan under grants 2016YFB0800603 and 2017YFB1200704, and the National Natural Science Foundation of China under grant No. 61662066.

\section*{COMPLEMENTARY MATERIAL }

\textbf{Appendix A} Illustration of both simulation and analytical value relevant to the logarithm of diameters of graph models $N(p,q.4)$,  $N(p,q,9)$ and $N(0,1,9)$ as a function of parameter $p$.

\textbf{Appendix B} Illustration of three specified pearson correlation coefficients. They are in turn $r(0.1,0.9,t)$,  $r(0.5,0.5,t)$ and $r(0.9,0.1,t)$ as a function of time step $t$.

\begin{figure}
\centering
  \includegraphics[height=3cm]{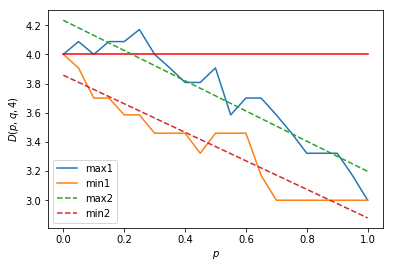}\quad
  \includegraphics[height=3cm]{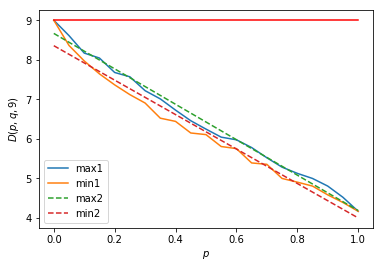}\\
{\small Fig.5. The diagrams of the logarithm of diameters of graph models $N(p,q,4)$, $N(p,q,9)$ and $N(0,1,9)$ as a function of parameter $p$. In both panels, solid lines represent simulation values and dotted lines are corresponding analytical ones. Among of them, blue lines may be obtained by taking average over sum of maximums when running our program exactly ten times. On the contrary, taking average over sum of minimums for orange lines. This red line is always $\ln |D(0,1,t)|$ corresponding to a specified time step $t$. Obviously, simulations are fitting for analytical values, indicating simulations are in strong agreement with our assertion.  }
\end{figure}

\begin{figure}
\centering
  \includegraphics[height=3cm]{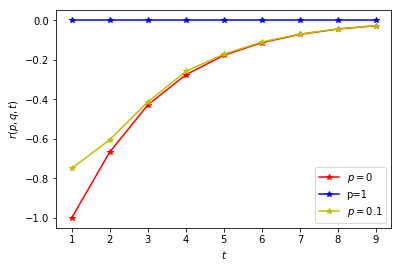}\quad
  \includegraphics[height=3cm]{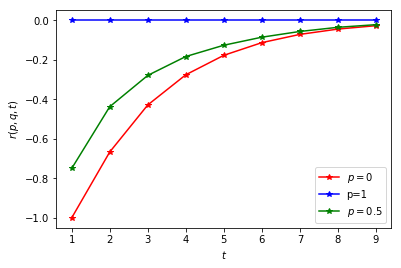}\quad
  \includegraphics[height=3cm]{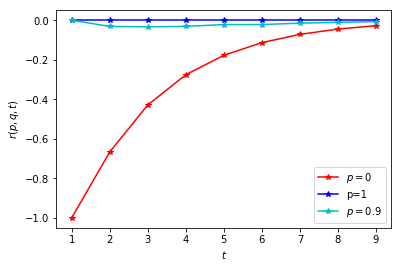}\\
{\small Fig.8. The diagrams of three specified pearson correlation coefficients, $r(0.1,0.9,t)$,  $r(0.5,0.5,t)$ and $r(0.9,0.1,t)$ at $1\leq t\leq9$.  }
\end{figure}

\end{document}